\documentclass[11pt]{article}

\usepackage{geometry,wasysym}
\usepackage{dsfont}

\usepackage{epsfig}

\begin{document}

\sffamily

\vspace*{1mm}

\begin{center}

{\LARGE Topology and index theorem with a \\ \vskip2mm generalized Villain lattice action -- a test in 2d}

\vskip10mm
Christof Gattringer, Pascal T\"orek 

\vskip5mm
Universit\"at Graz, Institut f\"ur Physik\footnote{Member of NAWI Graz.}, Universit\"atsplatz 5, 8010 Graz, Austria 
\end{center}
\vskip12mm

\begin{abstract}
Using 2-d U(1) lattice gauge theory we study two definitions of the topological charge constructed from a 
generalized Villain action and analyze the implementation of the index theorem based on the overlap Dirac 
operator. One of the two definitions expresses the topological charge as a sum of the Villain variables and 
treats charge conjugation symmetry exactly, making it particularly useful for studying related physics. Our 
numerical analysis establishes that for both topological charge definitions the index theorem becomes exact 
quickly towards the continuum limit. 
\end{abstract}

\vskip10mm

\noindent
{\bf \large Introduction}
\vskip3mm

\noindent
Topological terms are an important ingredient for many quantum field theories in high energy and condensed 
matter physics. The corresponding physical phenomena are non-perturbative in nature and suitable 
non-perturbative approaches are needed for properly describing them. An appealing approach is the 
lattice formulation, which, however, faces two key challenges when dealing with topological terms: 
1) The topological charge must be suitably discretized such that the symmetries one wants to study 
are correctly implemented on the lattice. 2) Topological terms generate a complex action 
problem that must be overcome for Monte Carlo simulations. 

Recently \cite{Sulejmanpasic:2019ytl} a new discretization approach for abelian topological terms was presented 
that gives rise to a generalized Villain action \cite{Villain:1974ir}, which now also includes the topological term. 
For many cases the complex action problem can be solved in this formulation by exactly mapping the system to dual
variables such that worm algorithms \cite{worm} and their generalization to gauge surfaces \cite{Mercado:2013yta} 
can be used for numerical simulations. In an application of this new formulation the 2-d U(1) gauge Higgs model at
topological angle $\theta = \pi$ was studied in \cite{Gattringer:2018dlw,Goschl:2018uma} (for earlier related  
studies see \cite{Wiese:1988qz,Kloiber:2014dfa,Gattringer:2015baa}). With the solution of the complex action problem 
by dualization and the correct implementation of charge conjugation symmetry at $\theta = \pi$ 
it was possible to study the spontaneous breaking of the charge conjugation symmetry and 
show \cite{Gattringer:2018dlw,Goschl:2018uma}
that the transition is in the 2-d Ising universality class as expected 
\cite{Affleck,Nahum:2011kq,Komargodski:2017dmc}

In this paper we further explore the formulation of topological terms based on \cite{Sulejmanpasic:2019ytl}, now focussing
on the index theorem \cite{Atiyah}. The index theorem connects the topological charge with the number of zero modes of
the Dirac operator and plays an important role in our understanding of the effects of gauge field topology on fermions.
Consequently a proper lattice discretization must implement the index theorem in a suitable way. We explore this 
property for the topological charge \cite{Sulejmanpasic:2019ytl} in 2 dimensions.

More specifically we use the overlap lattice Dirac operator \cite{overlap1,overlap2} for the index theorem. 
For the gauge fields we define two different forms of the topological charge, both based on the generalized Villain 
action \cite{Sulejmanpasic:2019ytl}. Using quenched ensembles on different volumes and at different gauge 
couplings we compute the index from the Dirac operator. We show that as one approaches the continuum limit 
the index theorem is obeyed perfectly for both variants of the topological charge. The study establishes that the 
topological charge based on the generalized Villain action \cite{Sulejmanpasic:2019ytl} implements the relevant
symmetries exactly and allows one to use the index theorem for exploring the physics of fermions 
coupled to gauge fields with topological terms.

\vskip6mm
\noindent
{\bf \large General Villain action, topological charge, index theorem}
\vskip2mm

\noindent
For coupling the lattice gauge fields to the fermions we use the compact gauge fields $U_{x,\mu} \in$ U(1) assigned to 
the links of a $N \times N$ lattice $\Lambda$ where they obey periodic boundary conditions. 
We parameterize the U(1)-valued compact gauge links with $\mathds{R}$-valued gauge fields $A_{x,\mu}$,
\begin{equation}
U_{x,\mu} \; = \; e^{\, i A_{x,\mu}} \; \; \; \; \mbox{with} \; \; \; \; A_{x,\mu} \, \in \; [-\pi, \pi) \; .
\label{Udef}
\end{equation}
For studying the index theorem we use the overlap Dirac operator $D^{ov}$. The overlap operator \cite{overlap1,overlap2}
is constructed from the Wilson Dirac operator $D^w$ given by (the lattice spacing is set to $a = 1$)
\begin{equation}
D^w_{\;x,y} \; = \; 2 \, \mathds{1}  \, \delta_{x,y} \, - \,
\sum_{\mu = 1}^2 \left[ \frac{\mathds{1} \!-\! \sigma_\mu}{2} \, U_{x,\mu} \, \delta_{x+\hat{\mu},y}
+ \frac{\mathds{1} \!+\! \sigma_\mu}{2} \, U_{y,\mu}^{\, *} \, \delta_{x-\hat{\mu},y} \right].
\label{DW}
\end{equation}
$\sigma_\mu$ are the Pauli matrices, $\mathds{1}$ the corresponding $2\times2$ unit matrix 
and for the fermions we use periodic boundary conditions in space and anti-periodic boundary conditions in time. 
Note that the mass parameter was set to 0 in $D^w$. The overlap operator $D^{ov}$ then is defined as
(here $\mathds{1}$ is the $2 N^2 \times 2 N^2$ unit matrix and $\gamma_5 \equiv \sigma_3$)
\begin{equation}
D^{ov} \; = \; \mathds{1} + A(\gamma_5 A \gamma_5 A)^{-1/2} \; \; \mbox{with} \; \; A = D^w - \mathds{1}(1+s) \; ,
\label{DOV}
\end{equation} 
where $s$ is a real parameter restricted to $|s| < 1$, which we here set to $s = 0.01$.

Using the overlap operator one can write the fermionic definition of the topological charge as \cite{indexlattice}
\begin{equation}
Q_F \; \equiv \; \frac{1}{2} \mbox{Tr} \big[ \gamma_5 D^{ov} \big] \; = \; n_- \, - \, n_+ \; .
\label{QF}
\end{equation}
The second equality is a trivial identity for any lattice Dirac operator that obeys the Ginsparg-Wilson relation
and relates $Q_F$ to the difference of the numbers
of left-handed ($n_-$) and right-handed ($n_+$) zero modes\footnote{We remark, that for 2-d U(1) gauge fields 
one can show explicitly that only one of the two numbers $n_-$, $n_+$ is non-zero. 
See, e.g., \cite{vanishing1,vanishing2,vanishing3} for this 
so-called ''vanishing theorem''.}. In other words, the fermionic definition of the 
topological charge simply expresses the index theorem. 

For defining the gauge field action we now use the real-valued gauge fields $A_{x,\mu}$. We define the lattice field
strength tensor as
\begin{equation}
F_{x,12} \; = \; A_{x+\hat1,2} \, - \, A_{x,2} \, - \, A_{x+\hat2,1} \, + \, A_{x,1} \; \equiv \; (dA)_{x,12} \; ,
\end{equation}
where in the last step we have defined the exterior derivative $(dA)_{x,12}$ on the lattice that constitutes the field strength.
$F_{x,12}$ is invariant under 
$A_{x,\mu} \rightarrow A_{x,\mu}^\prime = A_{x,\mu} - \lambda_{x+\hat \mu} + \lambda_{x}$, which together 
with $\psi_x \rightarrow e^{\, i \lambda_x} \psi_x$ for the fermion fields defines the U(1) gauge transformations.

Note, however, that using the compact link variables $U_{x,\mu}$ in the parameterization (\ref{Udef}) and in 
(\ref{DW}), (\ref{DOV}) implies a
second invariance of the Dirac operator and the fermion action, which is given by the shifts 
\begin{equation}
A_{x,\mu} \; \rightarrow \; A_{x,\mu} \; + \; 2\pi \, k_{x,\mu} \quad \mbox{with} \quad k_{x,\mu} \, \in \, \mathds{Z} \; .
\label{shift}
\end{equation}
Clearly this is not a symmetry of $F_{x,12}$ which transforms non-trivially under the shifts (\ref{shift}),
\begin{equation}
F_{x,12} \; \rightarrow \; F_{x,12} \; + \; 2\pi \, (dk)_{x,12} \; .
\label{Fshift}
\end{equation}
In order to take into account the symmetry under 
shifts, the Wilson gauge action uses the $2\pi$-periodic functions $\cos(F_{x,12})$
for defining the gauge action and in a similar way one may construct the field theoretical discretization of the topological
charge as $Q = \sum_x \sin(F_{x,12}) / 2\pi$. However, this does not give rise to an integer valued definition of the 
topological charge, which is obtained only in the continuum limit. As a consequence this discretization also 
does not correctly implement the non-trivial charge conjugation symmetry at topological angle $\theta = \pi$. 

An alternative discretization provides the so-called Villain action. Invariance of the gauge action under the shifts 
(\ref{Fshift}) is obtained by introducing a new plaquette-based variable $n_{x} \in \mathds{Z}$ and replacing 
$F_x \equiv F_{x,12}$ by $F_x + 2 \pi n_x$.  Subsequently one sums $n_{x}$ over all integers such that the
action becomes a $2\pi$-periodic function and thus invariant under the shifts (\ref{Fshift}). Actually one may 
consider $n_x$ as new (additional) plaquette-valued gauge field that ensures invariance under the
local shift transformation (see \cite{Sulejmanpasic:2019ytl,Gaiotto:2017yup} for detailed discussions of these aspects).

This construction is now used for the discretization of both, the gauge action $S_G = 1/2e^2 \! \int \! d^2x F_{12}(x)^2,$
as well as the topological charge $Q = 1/2\pi \int d^2x F_{12}(x)$. In this way we obtain the generalized Villain 
Boltzmann factor with topological term, 
\begin{equation}
B[A] \; = \; \prod_x \sum_{n_x \in \mathds{Z}} \, e^{ \, - \, \frac{\beta}{2} (F_x + 2 \pi n_x)^2}  \,
e^{ \, - \, i \frac{\theta}{2\pi} (F_x + 2 \pi n_x)} \; .
\end{equation} 
The product is over all plaquettes of the lattice, which in two dimensions can be labeled by the coordinate $x$ 
of their lower left corner. 

We introduce the  notation $\sum_{\{ n \} } \equiv \prod_x \sum_{n_x \in \mathds{z}}$ to write the 
Boltzmann factor as
\begin{equation}
B[A] \; = \;  \sum_{\{ n \} }  \, e^{ \, - \, \frac{\beta}{2} \sum_x (F_x + 2 \pi n_x)^2}  \,
e^{ \, - \, i \frac{\theta}{2\pi} \sum_x (F_x + 2 \pi n_x)} 
\;  = \;   \sum_{\{ n \} }  \, e^{ \, - \, \frac{\beta}{2} \sum_x (F_x + 2 \pi n_x)^2}  \,
e^{ \, - \, i \theta \sum_x n_x} \; .
\label{BAfinal}
\end{equation} 
In the second step we have made use of the fact that $\sum_x F_x = \sum_x (dA)_x = 0$ for a lattice 
with periodic boundary conditions. As a consequence we may identify the topological charge as the factor that
multiplies $- i \theta$ in the exponent and obtain 
\begin{equation}
Q_V \; = \; \sum_x n_x \; .
\label{QVdef}
\end{equation}
Note that this is a definition in terms of only the Villain variables $n_x$ and we thus refer to $Q_V$ 
as the ''Villain form of the topological charge''. 

We stress that $Q_V$ defined in (\ref{QVdef}) is an integer and as a consequence also the non-trivial 
charge conjugation at $\theta = \pi$ is implemented exactly. More specifically, for the gauge field charge conjugation
corresponds to $A_{x,\mu} \rightarrow - A_{x,\mu}$. This implies $U_{x,\mu} \rightarrow U_{x,\mu}^\star$, which in turn 
gives rise to the usual charge conjugation invariance of the fermion action based on the Wilson- or overlap Dirac operators
in Eqs.~(\ref{DW}) and (\ref{DOV}). To see the invariance of the Boltzmann factor $B[A]$ we note that under 
charge conjugation the field strength transforms as $F_x \rightarrow - F_x$ and that in the gauge field action 
(the quadratic term in the exponent of (\ref{BAfinal})) this change of sign 
is compensated by transforming the Villain variables 
as $n_x \rightarrow - n_x$. This in turn implies $Q_V \rightarrow - Q_V$ and the part of the Boltzmann factor
that contains the topological charge, i.e., $e^{-i \theta Q_V}$, is invariant under $Q_V \rightarrow - Q_V$ for $\theta = \pi$
(and of course the trivial value $\theta = 0$). We thus see that the Villain definition of the topological charge exactly
implements the non-trivial charge conjugation symmetry at $\theta = \pi$.

As we have discussed, the Villain form of the topological charge $Q_V$ is defined in terms of the Villain variables $n_x$.
It is also possible to sum the Villain variables completely in the Boltzmann factor, such that 
$B[A]$ can then be written as a product over all plaquettes and each factor will still depend on the topological angle 
$\theta$. For this variant we may define a different form of 
the topological charge via a derivative of $\ln B[A]$ with respect to $\theta$,
\begin{eqnarray}
Q_S & \!\! \equiv \!\! & i \frac{\partial}{\partial \theta} \ln B[A] \, \Bigg|_{\theta = 0}= \; \frac{1}{B[A]} \; 
i \frac{\partial}{\partial \theta} 
\left( \prod_x \sum_{n_x \in \mathds{Z}} e^{ - \, \frac{\beta}{2} (F_x + 2 \pi n_x)^2 \, - \, i \theta n_x} \right)
\! \Bigg|_{\theta = 0}
\label{QSdef} \\
& \!\! = \!\! & \frac{1}{B[A]}  \, \sum_{y} \! \left(\prod_{x \neq y} \sum_{n_x \in \mathds{Z}} 
e^{ - \, \frac{\beta}{2} (F_x + 2 \pi n_x)^2}\!\right) \!\!
\sum_{n_{y} \in \mathds{Z}} \! n_{y} \; 
e^{ - \, \frac{\beta}{2} (F_{y} + 2 \pi n_{y})^2} 
\;  = \; \sum_y
\frac{\sum_{n_{y} \in \mathds{Z}}  \, n_y \; e^{ - \, \frac{\beta}{2} (F_{y} + 2 \pi n_{y})^2 } }
{\sum_{n_{y} \in \mathds{Z}}  e^{ - \, \frac{\beta}{2} (F_{y} + 2 \pi n_{y})^2} } \; .
\nonumber 
\end{eqnarray}
In this version of the topological charge the Villain variables are summed and $Q_S$ depends only on the gauge fields 
$A_{x,\mu}$ that enter via $F_x$. We refer to $Q_S$ as the "summed form of the topological charge". It is obvious that 
$Q_S$ is odd under charge conjugation ($F_x \rightarrow -F_x$ and $n_x \rightarrow -n_x$), but clearly 
$Q_S$ is not integer. We will see, however, that in the continuum limit $Q_S$ becomes concentrated on integers. 

\vskip6mm
\noindent
{\bf \large Numerical results for topological charge and index theorem}
\vskip2mm

\noindent
We now explore the properties of the topological charge definitions $Q_V$ and $Q_S$ based on the generalized 
Villain action and analyze their correlation with the fermionic definition $Q_F$ to test the index theorem. We use 
ensembles of quenched gauge field configurations generated with the Villain action on lattices with sizes 
between $V = 8^2$ up to $V = 24^2$ with statistics of $10^5$ to $2 \times 10^5$ configurations. 
We use a local algorithm that
updates the gauge field variables $A_{x,\mu}$ and the Villain variables $n_x$ in turn. This is combined with an 
additional update step 
where we shift $A_{x,\mu}$ by $\pm 2 \pi$ and the Villain variables $n_x$ and $n_{x-\hat{\nu}}, \nu \neq \mu$  
connected to $A_{x,\mu}$ by $\pm 1$ and $\mp 1$ accordingly, 
whenever the Monte Carlo proposal for $A_{x,\mu}$ is outside the fundamental domain $[-\pi, \pi)$.
The approach with updating the Villain variables $n_x$ as independent variables makes them 
explicitly available for computing the Villain form $Q_V$ of the topological charge\footnote{We remark that one can 
also use an update of only the gauge fields $A_{x,\mu}$ that for each acceptance state locally sums the 
Villain variables $n_x$ for the plaquettes containing $A_{x,\mu}$.}.
We study the approach to the continuum limit with a fixed physical volume by sending $\beta \rightarrow \infty$ 
at constant ratio $R = V/\beta$ using three values for that dimensionless ratio, $R = 32$, $R=64$ and $R = 128$, 
which correspond to continuum limits at different physical volumes. 

The strategy for calculating the fermionic definition $Q_F$ of the 
topological charge (\ref{QF}) is as follows: We solve the eigenvalue problem for the Hermitian matrix $\gamma_5 A$ 
explicitly and compute $D^{ov}$ in (\ref{DOV}) by employing the 
spectral theorem. The result is plugged into (\ref{DOV}) and the eigenvalues are computed again explicitly for 
evaluating the trace in (\ref{QF}), where the multiplication with $\gamma_5$ under the trace 
just weights the eigenvalues with $\pm 1$.

\begin{figure}[t!] 
  \vspace{-5mm}
  \begin{center}
   \hspace*{-4mm}
   \includegraphics[width=80mm]{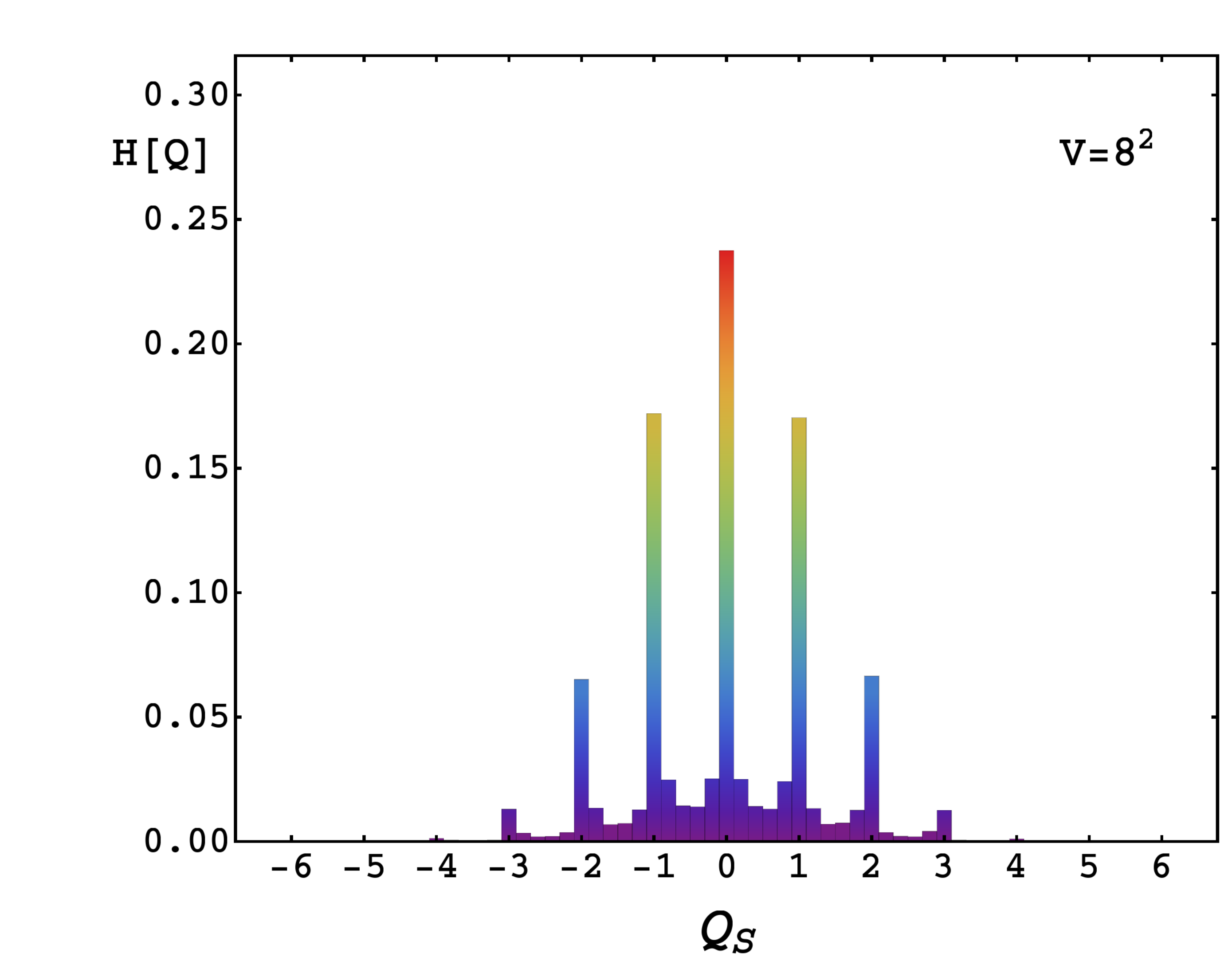} \hspace{-1mm}
   \includegraphics[width=80mm]{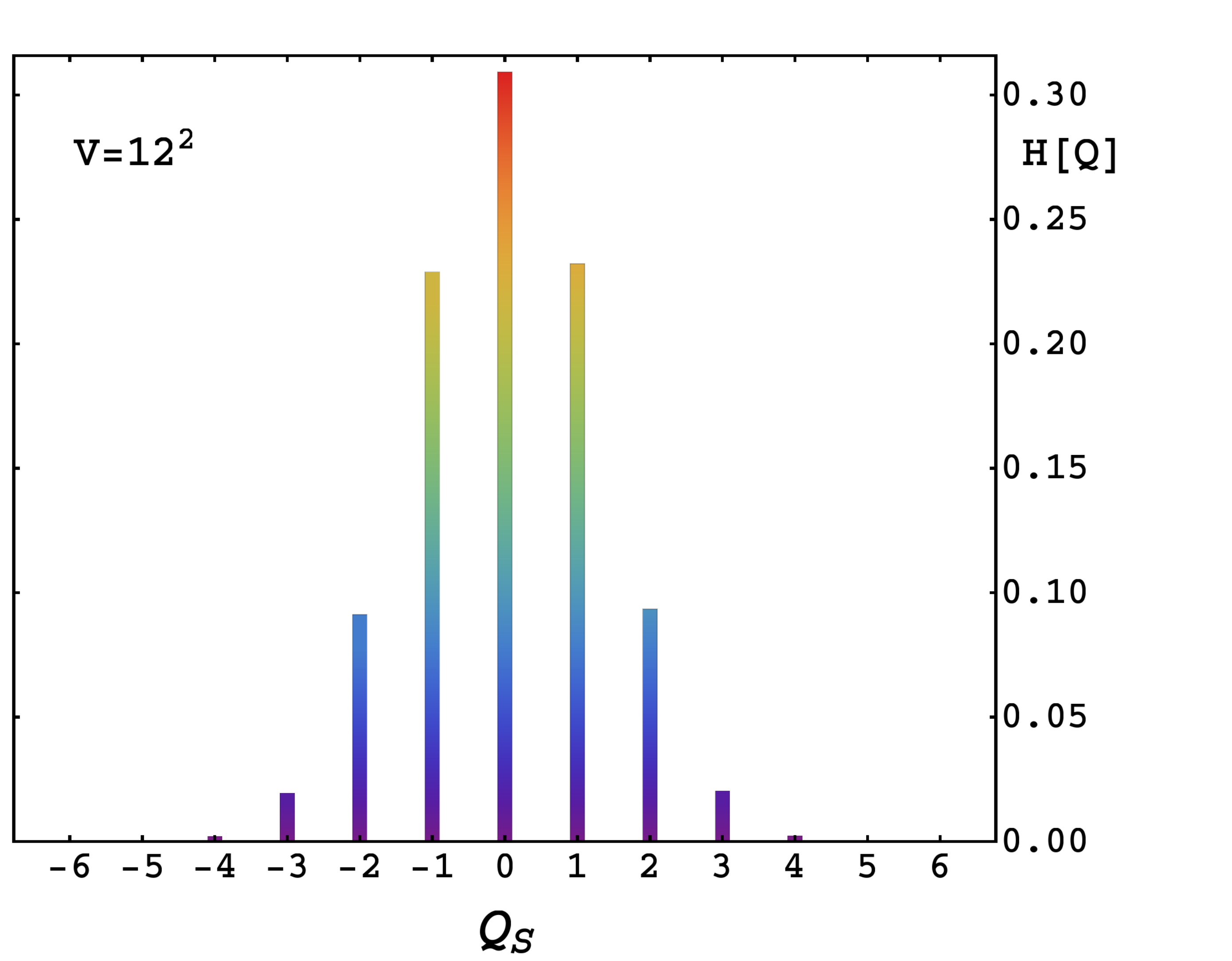}    
   \end{center}
   \vspace{-5mm}
   \caption{Histogram for the distribution of the topological charge $Q_S$ for two ensembles from the $R = 64$ set. 
   In the lhs.~plot we show the normalized histogram for $8 \times 8$ at $\beta = 1.0$ and on the rhs.~for $12 \times 12$ at 
   $\beta = 2.25$.} 
   \label{fig:QS_histo}
\end{figure}

Before we come to comparing the different topological charge densities and checking the index theorem, we first 
need to test whether the summed topological charge $Q_S$ defined in (\ref{QSdef}) becomes integer when 
approaching the continuum limit as expected. Indeed we find that $Q_S$ very quickly becomes concentrated on 
integers, as we illustrate in Fig.~\ref{fig:QS_histo}, where we show normalized histograms for the distribution 
of the topological charge $Q_S$ for our $R = 64$ ensembles. The lhs.\ plot is for the smallest value $\beta = 1.0$ of the
inverse gauge coupling we considered, i.e., most remote from the continuum limit. Nevertheless already here we see very 
pronounced peaks of the distribution near the integers and only a small fraction of the configurations has values of $Q_S$
outside the bins around the integers. Already at the next value of the inverse gauge coupling we consider, 
$\beta = 2.25$, all configurations are concentrated around the integers and we conclude, that although $Q_S$ 
is not an integer per definition, towards the continuum limit the values of $Q_S$ very quickly become restricted 
near integers. This finding, which in Fig.~\ref{fig:QS_histo} we illustrate for $R = 64$, was also confirmed 
for the other ratios $R = 32$ and $R = 128$. 
 
\begin{figure}[p] 
   \vspace*{-10mm}
   \begin{center}
   \includegraphics[width=73mm]{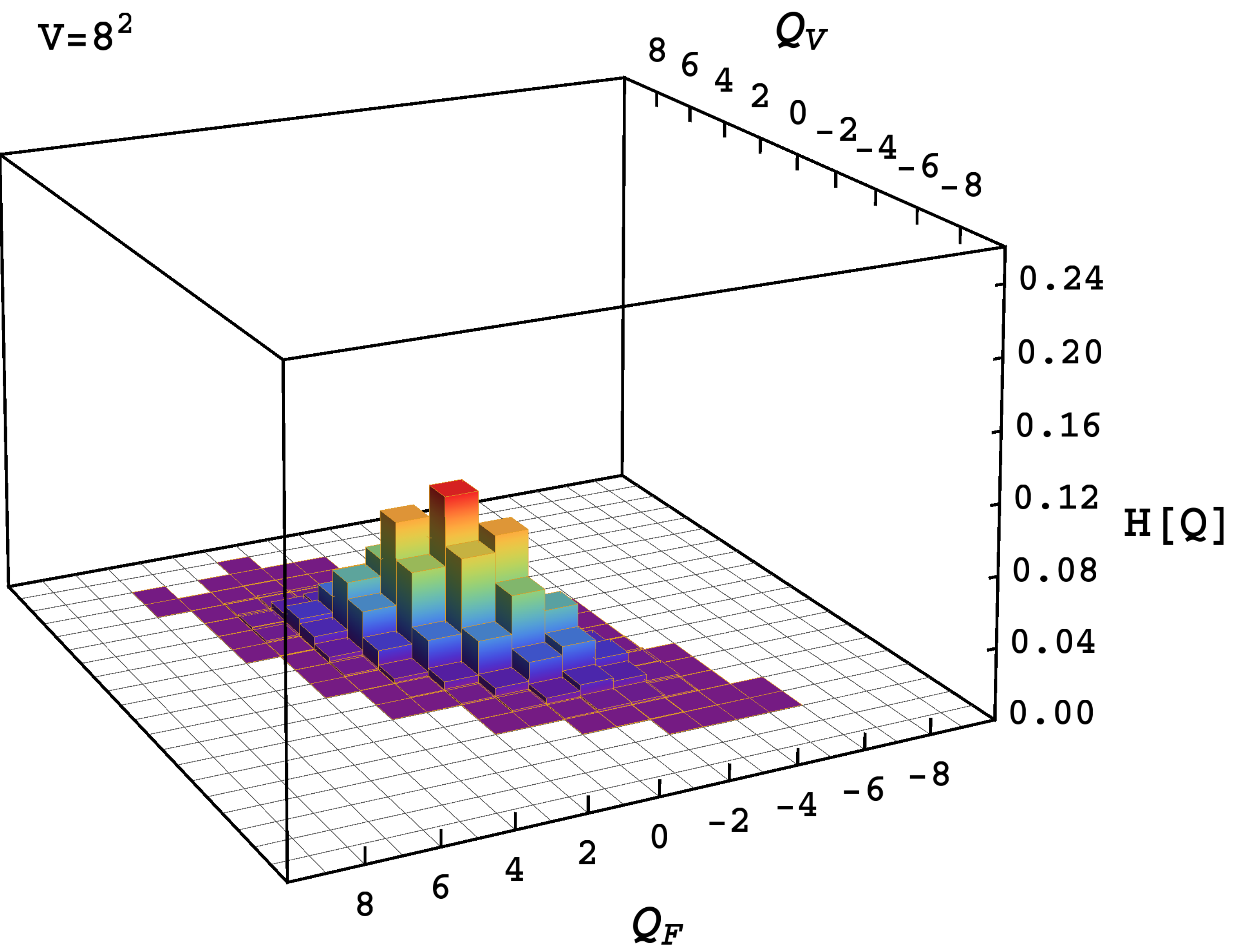} \hspace{2mm}
   \includegraphics[width=73mm]{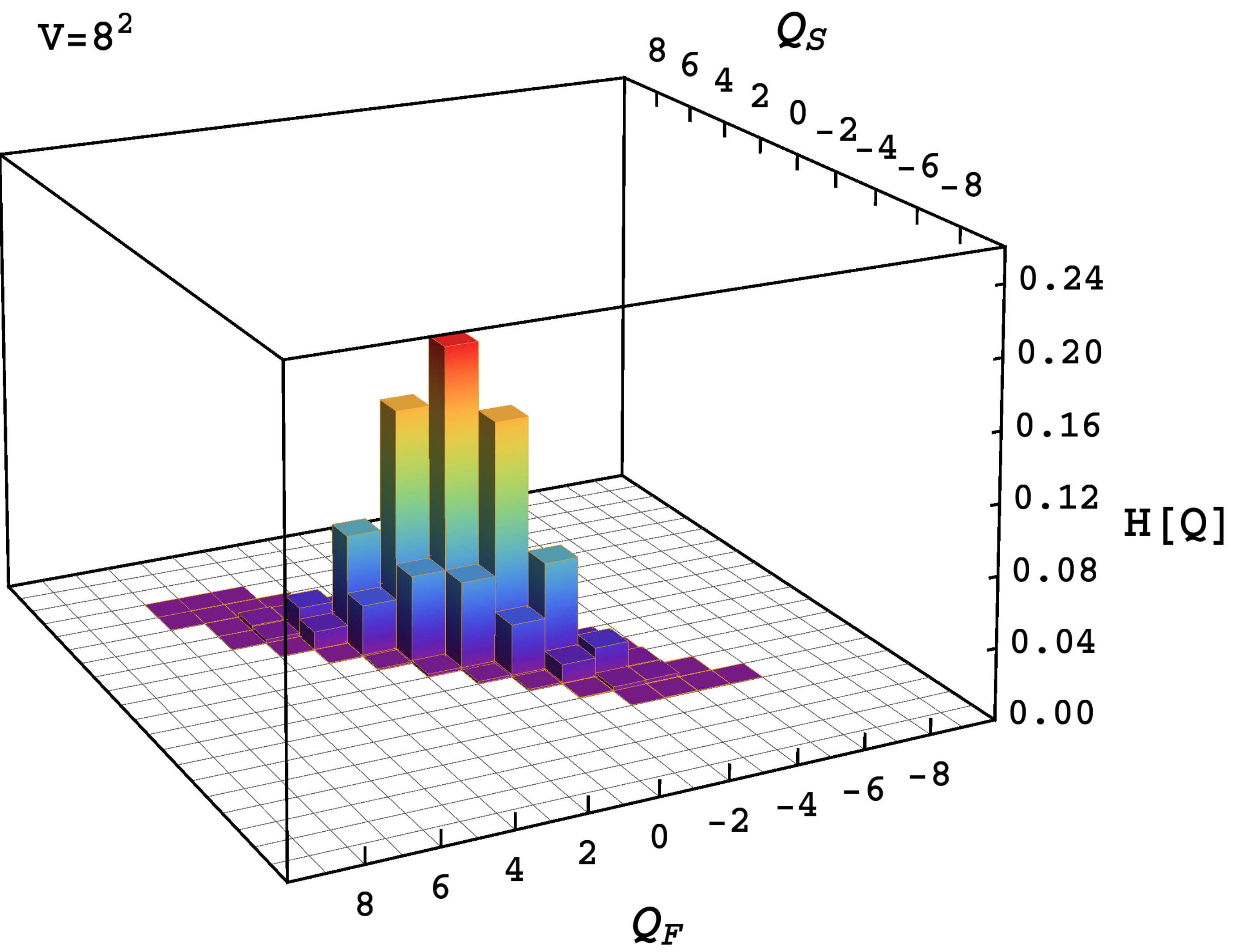}    
   \includegraphics[width=73mm]{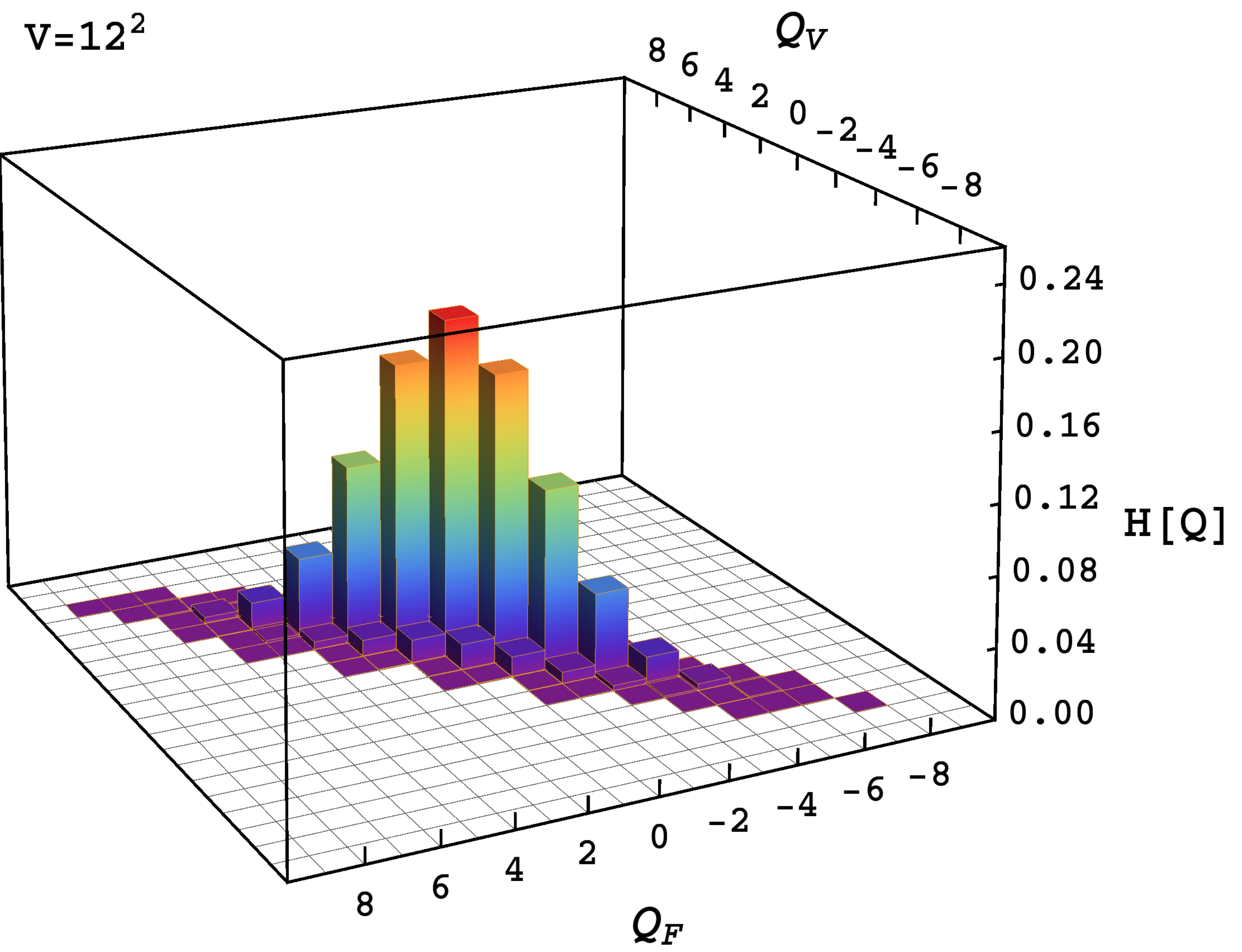} \hspace{2mm}
   \includegraphics[width=73mm]{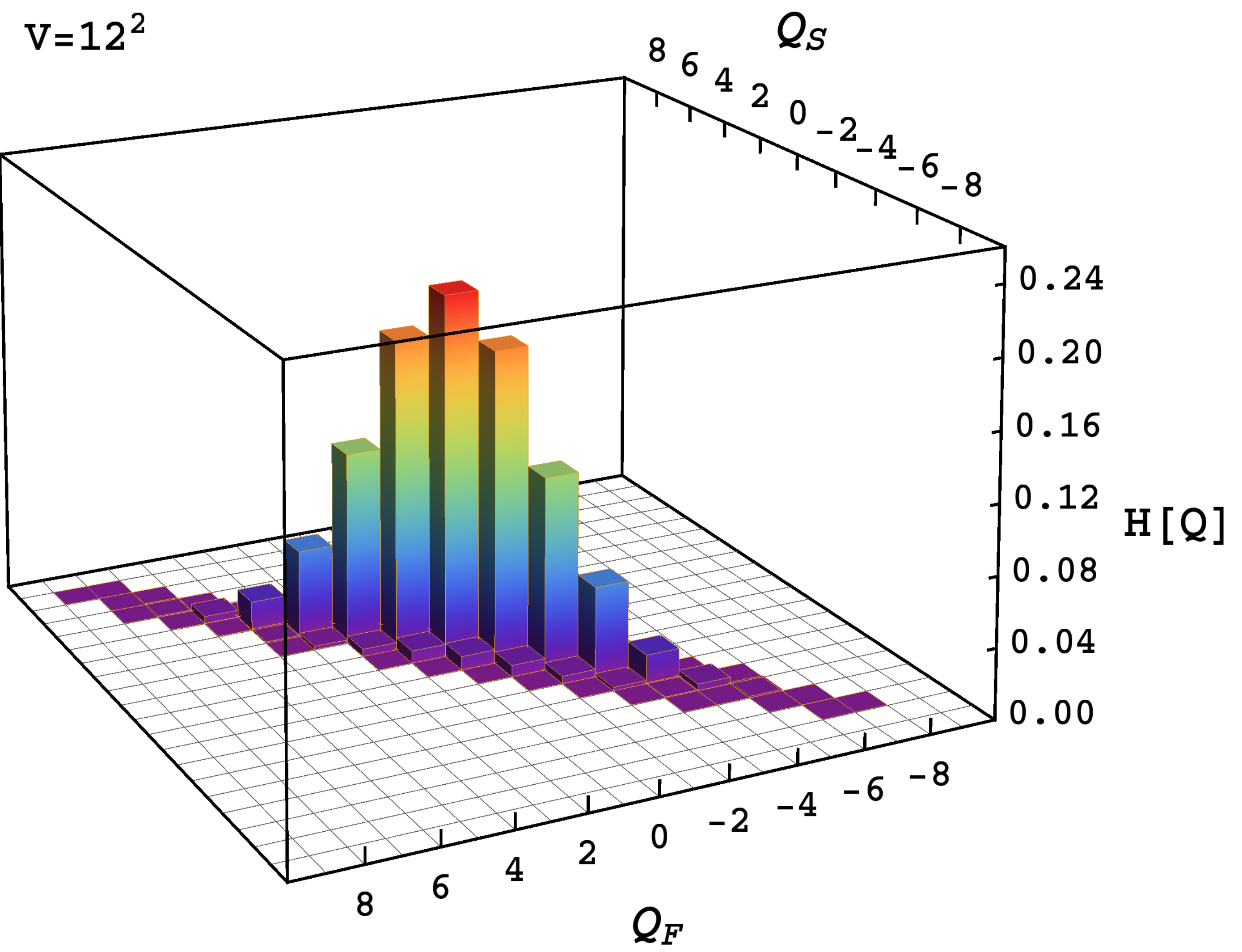}    
   \includegraphics[width=73mm]{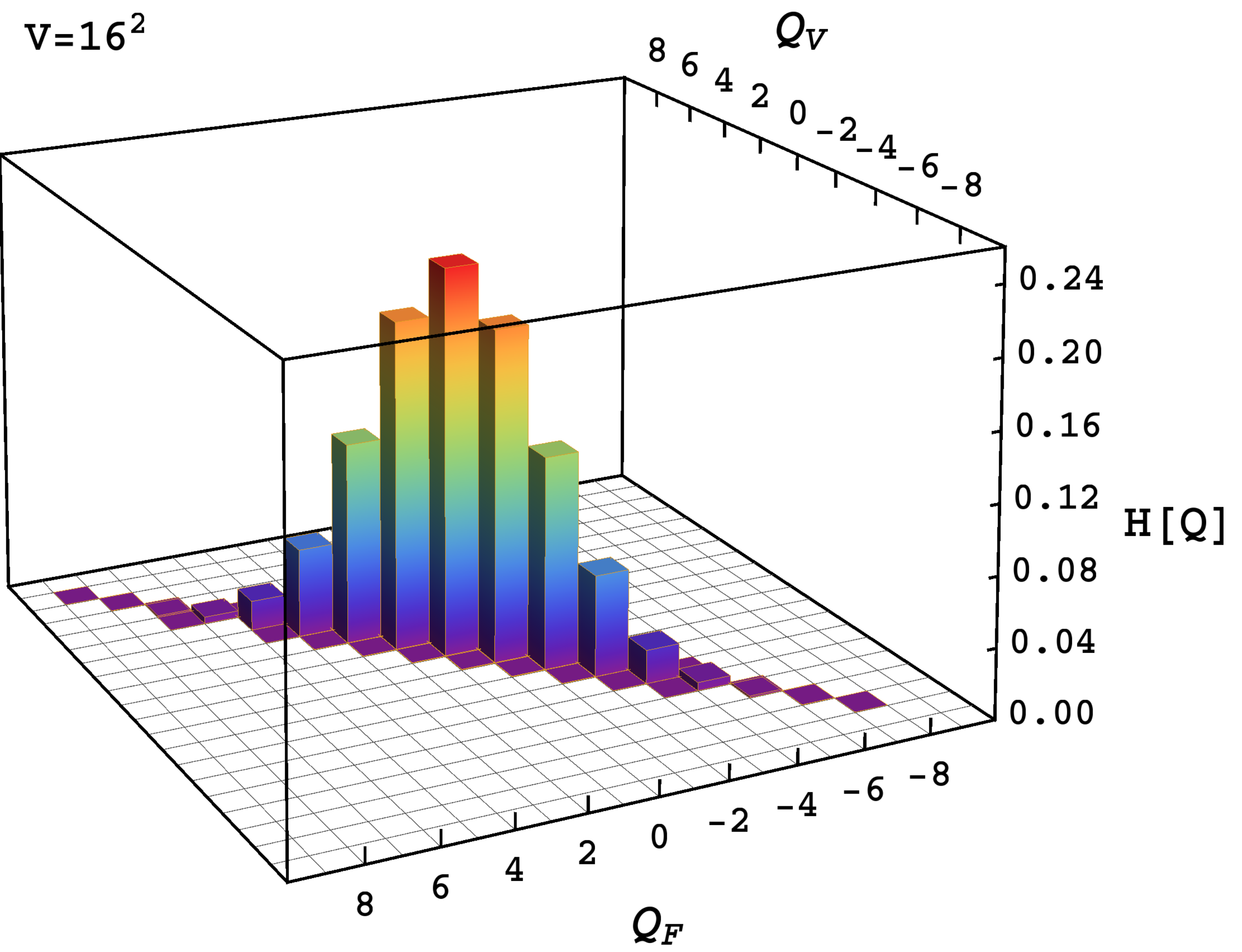} \hspace{2mm}
   \includegraphics[width=73mm]{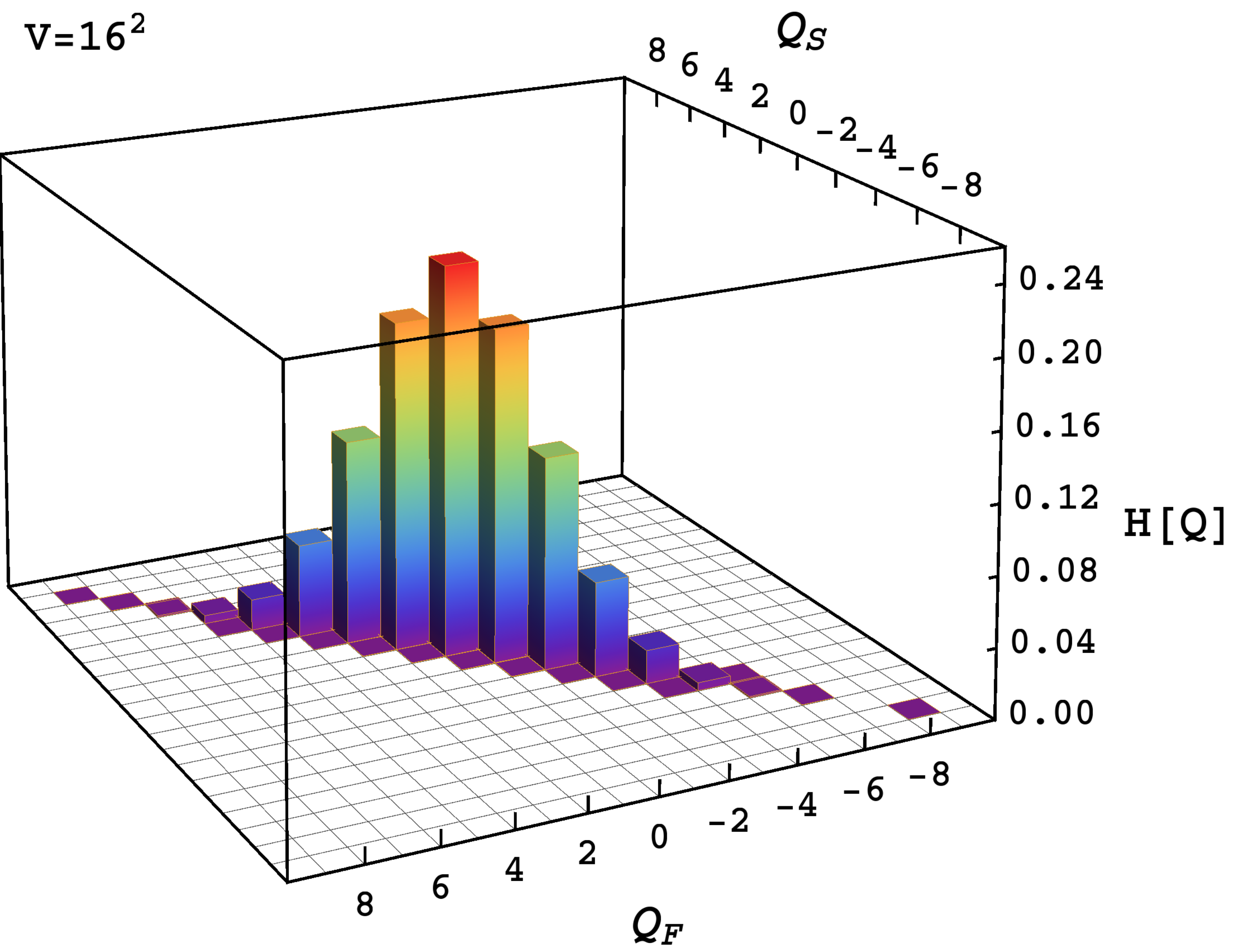}    
   \end{center}
   \caption{Histograms for the correlation of the topological charge definitions $Q_V$ with $Q_F$ (lhs.~column of plots) 
   and the correlation of $Q_S$ with $Q_F$ (rhs.~column of plots). In all plots $Q_F$ runs on the axis from right to left,
   and $Q_V$ (lhs.~column) and $Q_S$ respectively (rhs.) run from the front to the back. For each configuration we
   determined $Q_V, Q_S$ and $Q_F$ and counted how many configurations fall into the intervals for the 
   respective pair of values of $Q$. Subsequently we normalized the histograms such that the sum over 
   all contributions equals to 1. The results shown are for
   the $R = 128$ ensemble with (top to bottom) $8 \times 8, \beta = 0.5$, $12 \times 12, \beta = 1.125$ and 
   $16 \times 16, \beta = 2.0$.} 
   \label{fig:Q_comp}
\end{figure}

The next part of our numerical analysis of the Villain topological charge $Q_V$ and the summed topological charge 
$Q_S$ is to test how well the index theorem is obeyed and how this relation between topological charge and the number
of zero modes behaves towards the continuum limit. To assess these questions we begin with the sequence of histograms
shown in Fig.~\ref{fig:Q_comp}. For each configuration in our ensembles we determined $Q_V$, $Q_S$ and $Q_F$ and
counted the entries that fall into bins in the $Q_V$-$Q_F$ plane (for the lhs.\ column of plots) and in the 
$Q_S$-$Q_F$ plane (for the rhs.\ column of plots). Subsequently the histograms were normalized to 1. In all histogram
plots $Q_F$ runs on the axis from right to left, and $Q_V$ (lhs.~column) and $Q_S$ respectively (rhs.) 
from the front to the back. The histograms we show are for the $R = 128$ ensembles with 
$\beta = 0.5, V = 8^2$ in the top row, $\beta = 1.25, V = 12^2$ in the middle, and $\beta = 2.0, V = 12^2$ in the bottom 
row, i.e., we approach the continuum limit from top to bottom.

It is obvious that all histograms in Fig.~\ref{fig:Q_comp} show a strong correlation of the respective Villain-based 
definitions $Q_V$ (lhs.) and $Q_S$ (rhs.) with the fermionic definition $Q_F$ of the topological charge based on 
the index of the overlap operator. Furthermore both definitions converge towards a perfect correlation, i.e., 
non-zero histogram entries only on the diagonal, as we approach the continuum limit (top to bottom). Note that the 
fact that $Q_V$ and $Q_S$ both become perfectly correlated with $Q_F$ towards the continuum limit, of course
implies that also $Q_V$ and $Q_S$ become perfectly correlated in that limit.   

\begin{figure}[t] 
  \vspace*{-5mm}
  \begin{center}
   \includegraphics[width=80mm,clip]{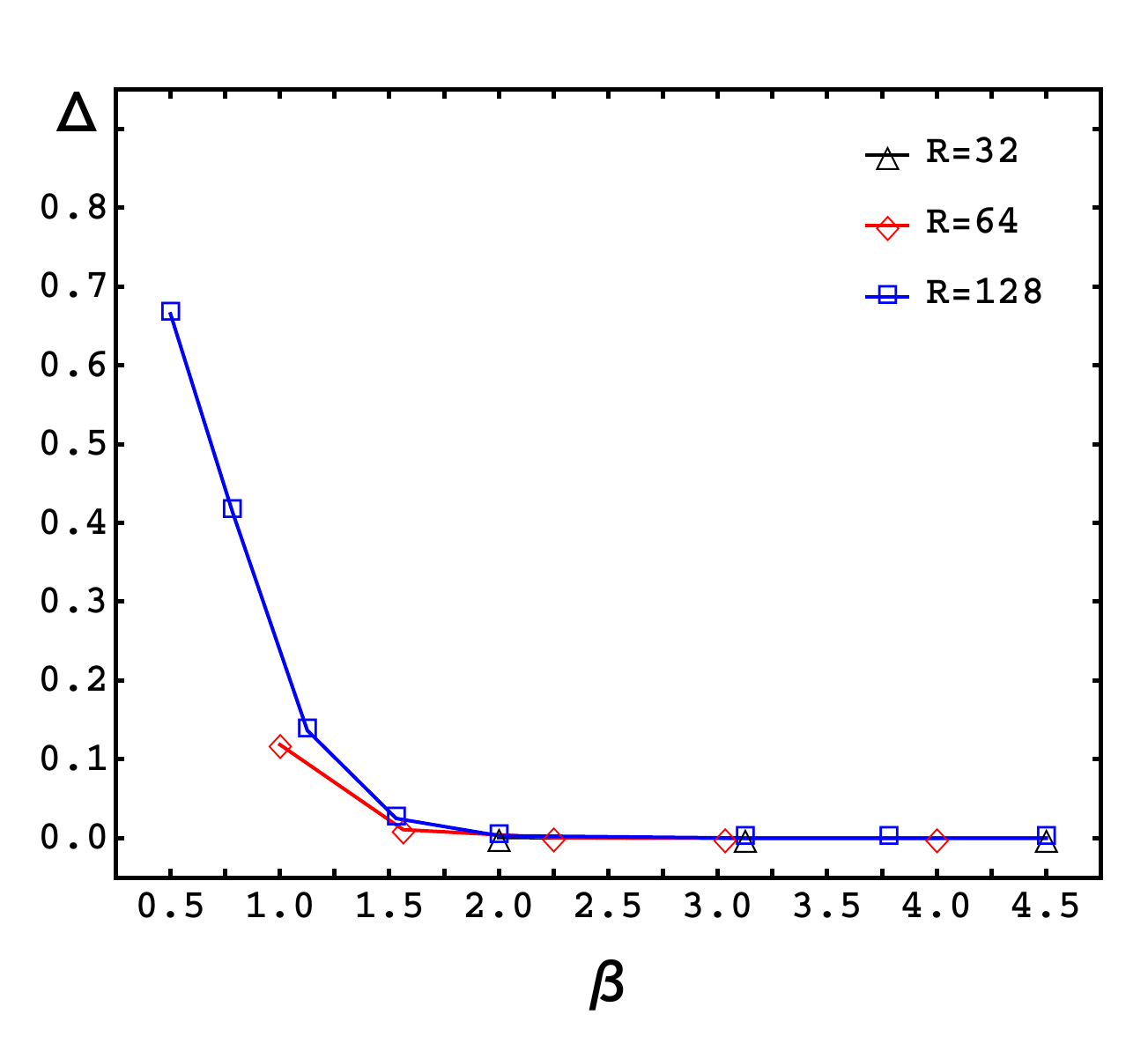} 
   \hspace{-4mm}
   \includegraphics[width=80mm,clip]{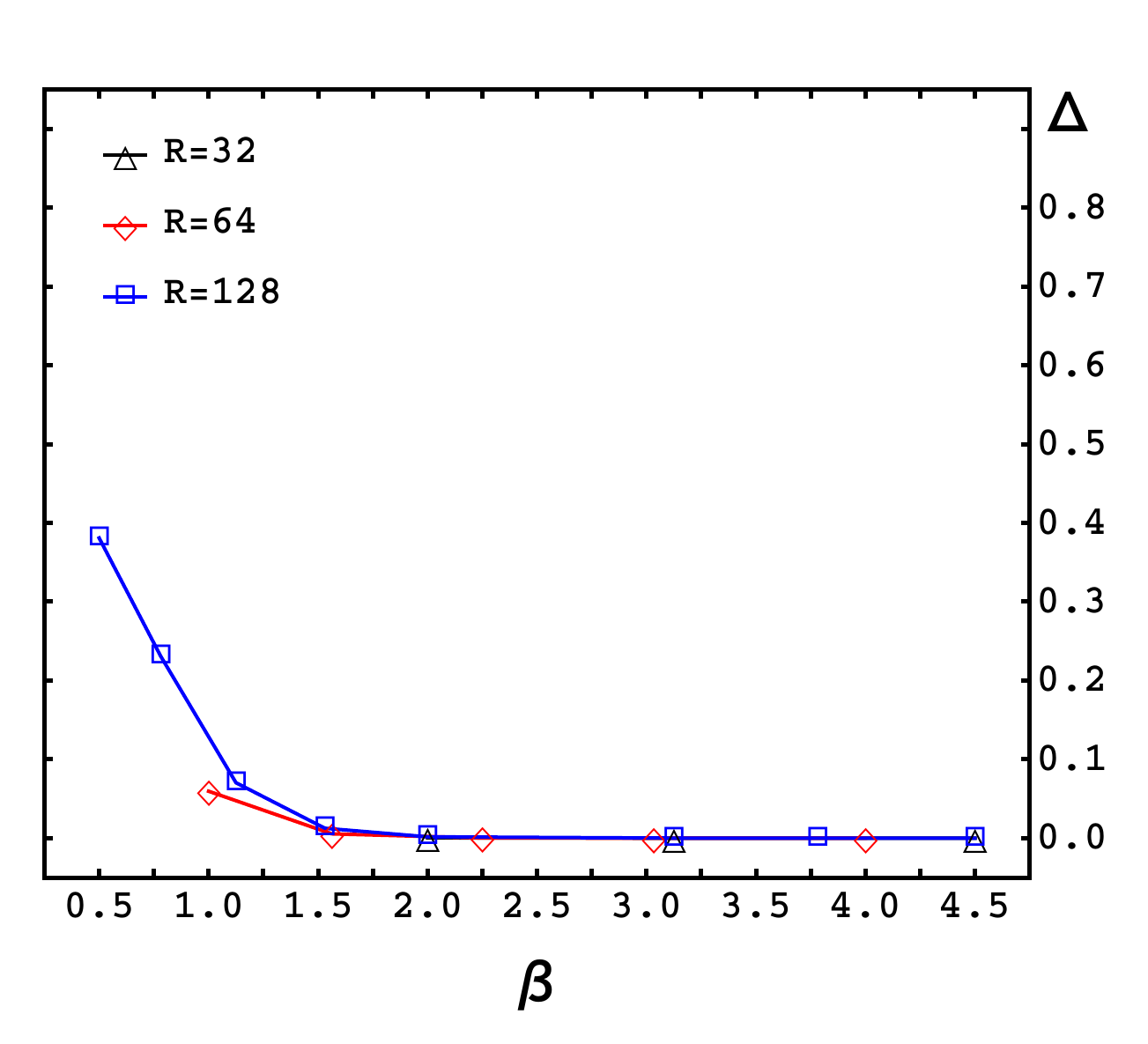}  
   \end{center}  
   \caption{Relative mismatch of $Q_V$ (lhs.\ plot) and $Q_S$ (rhs.) compared to the fermionic definition $Q_F$. 
   We show the results for all ensembles, i.e., all values of $R = V/\beta$ and all volumes $V$, and plot the mismatch as a 
   function of $\beta$.} 
   \label{fig:mismatch}
\end{figure}

It is important to understand that the two definitions $Q_V$ and $Q_S$ are conceptually quite different and thus 
also have different properties which are partly reflected in the histograms shown in Fig.~\ref{fig:Q_comp}: $Q_S$ 
defined in (\ref{QSdef}) is directly expressed in terms of the gauge fields $A_{x,\mu}$ that enter also the Dirac 
operator via the link variables $U_{x,\mu}$ (note that the Villain variables $n_x$ are summed over in (\ref{QSdef})). 
On the other hand, $Q_V$ defined in Eq.~(\ref{QVdef}) depends only 
on the Villain variables $n_x$, and thus couples to the fermions only indirectly via the gauge fields 
$A_{x,\mu}$ whose dynamics is connected to the Villain variables $n_x$ in the Boltzmann factor $B[A]$ in 
Eq.~(\ref{BAfinal}). This indirect connection of $Q_V$ and the fermions is the reason why at smaller $\beta$ 
for the Villain form $Q_V$ of the topological charge the correlation with $Q_F$ is less pronounced (top of the 
lhs.~column of plots 
in Fig.~\ref{fig:Q_comp}). Towards the continuum limit this effect is washed out and $Q_V$ and $Q_S$ both become 
equally well correlated with $Q_F$ and the index theorem is obeyed.
 
We conclude this study with making the analysis of the index theorem more quantitative. In Fig.~\ref{fig:mismatch} we show 
as a function of the inverse gauge coupling $\beta$ the mismatch $\Delta$, defined as the fraction of configurations 
where $Q_V$ and $Q_F$ (lhs.\ plot), and $Q_S$ and $Q_F$ respectively (rhs.\ plot) disagree, i.e., the fraction
of configurations where the index theorem is violated. We show the data for all our ensembles, i.e., for the three 
dimensionless ratios $R = V/\beta$ and use all volumes up to $V = 24^2$. The results for $\Delta$ show that 
the fraction of configurations where the index theorem is violated drops quickly towards the continuum limit, and 
above $\beta \sim 2.5$ we see no more violations of the index theorem. This holds for both definitions of the topological
charge, $Q_V$ and $Q_S$ at all values of $R$. We conclude that in the continuum limit 
the index theorem holds for both topological charges based on the generalized Villain action. 

\vskip6mm
\noindent
{\bf \large Discussion}
\vskip2mm

\noindent
In this letter we have analyzed new definitions of the topological charge that are based on a generalized 2-d U(1) 
Villain action and have studied how the index theorem based on the overlap operator becomes manifest in the 
continuum limit. In particular the Villain form $Q_V$, where the topological charge is an integer, simply the sum 
of the Villain variables $n_x$, is interesting since it exactly implements the charge conjugation symmetry at 
$\theta = \pi$, which in \cite{Gattringer:2018dlw,Goschl:2018uma} was used to determine the critical exponents 
at the corresponding transition in the gauge Higgs model. 

When using the new formulation of the topological charge to study systems with fermions it is of 
foremost interest to establish that the
index theorem is obeyed since it provides the link between gauge field topology and the fermions. Since the Villain 
form $Q_V$ of the topological charge depends only on the Villain variables $n_x$, while the fermions couple
only to the gauge fields $A_{x,\mu}$, it is not a-priori clear how well the index theorem is implemented. Our study
shows that indeed the index theorem becomes essentially exact very quickly when going towards the continuum limit.
This holds for the continuum limits at different values of $R = V/\beta$, i.e., at different physical volumes. Also the 
second definition of the topological charge, the summed form $Q_S$ that depends only on the gauge fields, shows
a fast approach towards an exact index theorem in the continuum limit.

The study presented here is the analysis of the index theorem for the simplest case of abelian topological terms. 
In \cite{Sulejmanpasic:2019ytl} several cases of abelian topological 
terms (compare also \cite{Sulejmanpasic:2018upi,Tanizaki:2018xto}) based on the Villain action in 2-d, 3-d
and 4-d were worked out and it would be interesting to explore their relation to the index of the Dirac operator. 
These are issues to be analyzed in future work, with the current study establishing an encouraging example. 
 
\vskip5mm
\noindent
{\bf Acknowledgments:}
\vskip2mm
\noindent
We would like to thank Daniel G\"oschl and Tin Sulejmanpasic for discussions. 
This work is supported by the Austrian Science Fund FWF, grant I 2886-N27. 
Parts of the computational results presented have been achieved using the Vienna Scientific Cluster (VSC).
The authors gratefully acknowledge support from NAWI Graz.


\begin{thebibliography}{12}

\bibitem{Sulejmanpasic:2019ytl}
  T.~Sulejmanpasic, C.~Gattringer,
  Nucl. Phys. {\bf B 943C} (2019) 114616 [arXiv:1901.02637].

\bibitem{Villain:1974ir}
  J.~Villain,
  J.\ Phys.\ (France) {\bf 36} (1975) 581.

\bibitem{worm}
  N.~Prokof'ev, B.~Svistunov,
  Phys.\ Rev.\ Lett.\  {\bf 87} (2001) 160601.

\bibitem{Mercado:2013yta}
  Y.~Delgado Mercado, C.~Gattringer, A.~Schmidt,
  Comput.\ Phys.\ Commun.\  {\bf 184} (2013) 1535
  [arXiv:1211.3436].
  
\bibitem{Gattringer:2018dlw}
  C.~Gattringer, D.~G{\"o}schl, T.~Sulejmanpasic,
  Nucl.\ Phys.\  {\bf B 935} (2018) 344
  [arXiv:1807.07793].
  
\bibitem{Goschl:2018uma}
  D.~G{\"o}schl, C.~Gattringer, T.~Sulejmanpasic,
  PoS LATTICE2018 (2019) 226 [arXiv:1810.09671].
    
\bibitem{Wiese:1988qz}
  U.-J.~Wiese,
  Nucl.\ Phys.\ {\bf B 318} (1989) 153.

\bibitem{Kloiber:2014dfa}
  T.~Kloiber, C.~Gattringer,
  PoS LATTICE {\bf 2014} (2014) 345.

\bibitem{Gattringer:2015baa}
  C.~Gattringer, T.~Kloiber, M.~M{\"u}ller-Preussker,
  Phys.\ Rev.\  {\bf D  92} (2015)  114508
  [arXiv:1508.00681].
  
\bibitem{Affleck}  
I.~Affleck, Phys.\ Rev.\ Lett.\ {\bf 66} (1991) 2429.

\bibitem{Nahum:2011kq}
  A.~Nahum and J.~T.~Chalker,
  Phys.\ Rev.\ E {\bf 85} (2012) 031141
  [arXiv:1112.4818].

\bibitem{Komargodski:2017dmc}  
  Z.~Komargodski, A.~Sharon, R.~Thorngren, X.~Zhou,
  SciPost Phys.\  {\bf 6} (2019) 003
  [arXiv:1705.04786].
  
\bibitem{Atiyah}
   M.~Atiyah, I.M.~Singer, 
   Ann.\ Math.\ {\bf 93} (1971) 139.
   
\bibitem{overlap1}
H.~Neuberger, Phys.\ Lett.\ {\bf B 417} (1998) 141.

\bibitem{overlap2}
H.~Neuberger, Phys.\ Lett.\ {\bf B 427} (1998) 353.

\bibitem{indexlattice}
M.\ L{\"u}scher, P.\ Weisz, JHEP {\bf 0207} (2002) 049.
   
\bibitem{vanishing1}
J.\ Kiskis, Phys. Rev. {\bf D 15} (1977) 2329.

\bibitem{vanishing2}
N.K.\ Nielsen, B.\ Schroer, Nucl.\ Phys.\ {\bf B 127} (1977) 493.

\bibitem{vanishing3} 
M.M.\ Ansourian, Phys.\ Lett.\ {\bf 70 B} (1977) 301.

\bibitem{Gaiotto:2017yup}
  D.~Gaiotto, A.~Kapustin, Z.~Komargodski, N.~Seiberg,
  JHEP {\bf 1705} (2017) 091
  [arXiv:1703.00501].
  
\bibitem{Sulejmanpasic:2018upi}
  T.~Sulejmanpasic, Y.~Tanizaki,
  Phys.\ Rev.\ B {\bf 97} (2018) 144201
  [arXiv:1802.02153].
  
\bibitem{Tanizaki:2018xto}
  Y.~Tanizaki, T.~Sulejmanpasic,
  Phys.\ Rev.\ B {\bf 98} (2018) 115126
  [arXiv:1805.11423].
 
\end{thebibliography}
\end{document}